\documentclass[traditabstract]{aa}

\usepackage{natbib}
\usepackage{graphicx}
\usepackage{amssymb}
\usepackage{amsmath}
\usepackage[dvipsnames]{color}

\begin{document}

\newcommand{\3}{\ss}
\newcommand{\n}{\noindent}
\newcommand{\eps}{\varepsilon}
\newcommand{\be}{\begin{equation}}
\newcommand{\ee}{\end{equation}}
\newcommand{\bl}[1]{\mbox{\boldmath$ #1 $}}
\def\ba{\begin{eqnarray}}
\def\ea{\end{eqnarray}}
\def\de{\partial}
\def\msun{M_\odot}
\def\msol{M_\odot}
\def\te{T_{\rm eff}}
\def\logg{\log g}
\def\lmix{l_{\rm mix}}
\def\lint{l_{\rm mix}^{\rm int}}
\def\latm{l_{\rm mix}^{\rm atm}}
\def\Hp{H_{\rm P }}
\def\mv{M_{\rm V}}
\def\mi{M_{\rm I}}
\def\mj{M_{\rm J}}
\def\mk{M_{\rm K}}
\def\lsol{L_\odot}
\def\lbol{L_{\rm bol}}
\def\div{\nabla\cdot}
\def\grad{\nabla}
\def\rot{\nabla\times}
\def\ltsima{$\; \buildrel < \over \sim \;$}
\def\simlt{\lower.5ex\hbox{\ltsima}}
\def\gtsima{$\; \buildrel > \over \sim \;$}
\def\simgt{\lower.5ex\hbox{\gtsima}}

\newcommand{\cp}{\citep}
\newcommand{\ct}{\citet}
\newcommand{\cta}{\citetalias}

\title{New evolutionary models for pre-main sequence and main sequence low-mass stars down to the hydrogen-burning limit}
\titlerunning{New evolutionary models for pre-main sequence and main sequence  low-mass stars}

\author{Isabelle Baraffe \inst{1,2}, Derek Homeier \inst{2}, France Allard \inst{2}, and Gilles Chabrier \inst{2,1} }
\authorrunning{Baraffe et al.}

\offprints{I. Baraffe} 
   
\institute{
University of Exeter, Physics and Astronomy, EX4 4QL Exeter, UK
(\email{i.baraffe@ex.ac.uk})
\and
\'Ecole Normale Sup\'erieure, Lyon, CRAL (UMR CNRS 5574), Universit\'e de Lyon, France
(\email{derek.homeier@ens-lyon.fr, fallard@ens-lyon.fr, chabrier@ens-lyon.fr})
}

\date{}

\abstract{We present new models for low-mass stars down to the hydrogen-burning limit that consistently couple  atmosphere and interior structures, thereby  superseding the widely used BCAH98 models. The new models include updated molecular linelists and solar abundances, as well as atmospheric convection parameters calibrated on 2D/3D radiative hydrodynamics simulations. Comparison of these models with observations in various colour-magnitude diagrams for various ages shows significant improvement over previous generations of models. The new models can solve flaws that are present in the previous ones, such as the prediction of optical colours that are too blue compared to M dwarf observations. They can also reproduce the four components of the young quadruple system LkCa 3 in a colour-magnitude diagram with one single isochrone, in contrast to any presently existing model. In this paper we also highlight the need for consistency when comparing models and observations, with the necessity of using evolutionary models and colours based on the same atmospheric structures.
 }


\maketitle

\section{Introduction}

In 1998, our team released a set of evolutionary models for low-mass
stars \cp[][hereafter BCAH98]{Baraffe98} based on the so-called
NextGen atmosphere models \cp{Hauschildt99} that marked a new era of
models that consistently coupled  interior and atmosphere structures. These
models became very popular because they  successfully reproduce various observational constraints, such as mass-luminosity and mass-radius relationships and colour-magnitude diagrams.  The models, however,  had some important shortcomings, such as predicting optical ($V\!-\!I$) colours that are too blue for a given magnitude (see \S \ref{sect_optical}). Later generations of models included improved molecular linelists for various atmospheric absorbers, such as the AMES linelists used in the Dusty and Cond models \cp{Chabrier00b, Allard01, Baraffe03},  but they still show shortcomings (see \S \ref{sect_nearIR}).
After a long effort to solve these flaws, efforts have paid off with the release of current models that
supersede the BCAH98 models. In this paper, we describe the main physical ingredients of the models and compare them to a selection of observations that highlight the improvement of these new models over previous ones. 

\section{Model description}

Evolutionary calculations are based on the same input physics describing stellar and substellar interior structures as are used in \ct{Chabrier97} and \ct{Baraffe98}. The major changes concern the atmosphere models, which provide the outer boundary conditions for the interior structure calculation, and the colours and magnitudes for a given star mass at any given age. Substantial changes have been made since the NextGen atmosphere models used in the BCAH98 evolutionary models. A preliminary set of atmosphere models, referred to as the BT-Settl models \cp{Allard12, Allard12E,  Rajpurohit13}, include some of these changes,  
 which are briefly summarised below. More recent modifications concerning the treatment of convection are described in \S \ref{sect_convection}. %A summary of the main characteristics of previous and present atmosphere models is provided in Table \ref{table1}.
 
\subsection{Molecular linelists and cloud formation}
\label{sect_lines}
 Line opacities for several important molecules have been updated, notably 
 %linelists from the Exomol project \cp{Tennyson12}, namely 
 the water linelist from \ct{Barber06}, metal hydrides such as CaH,
 FeH, CrH, TiH from \ct{Bernath06}, 
  vanadium oxide from Plez (2004, {\em priv.\ comm.}), 
 and carbon dioxide from \ct{Tashkun04}. 
 %\ct{Allard12} had updated the strengths of the
   %$\alpha-\epsilon$ bands of titanium oxide in the AMES linelist
   %\cp{Schwenke98} using the revised oscillator strengths of
   %\ct{Hedgecock95}, but it has since emerged that this list itself
   %contains errors due to missing higher-level sub-bands (R.~Freedman,
   %{\em priv.\ comm}). 
   %For the present set of atmosphere models we
   %have replaced this list with the one from \ct{plezTiO}. 
   For TiO, the present set of atmosphere models uses the linelist from \ct{plezTiO}.
  This list is not as complete at high energies as the AMES linelist
  \cp{Schwenke98} adopted in \ct{Allard01,Allard12}, 
  with only 11$\times 10^6$ lines
   compared to the 160$\times 10^6$ of \ct{Schwenke98}. But the Plez linelist reproduces
   the overall band strengths
better and thus generally improves the
   optical colours (Fig.~\ref{fig_magviv}). Obviously, the field is
   still in need of a new, complete, and accurate theoretical TiO
   linelist to allow quantitative high-resolution spectroscopic
   analysis of this important molecule, as recently pointed by the high-resolution transmission spectrum analysis of a transiting exoplanet \cp{hoeijmakers14}.
%   -- {\em France, could you provide the Hedgecock95 reference?}}
 
%\subsection{Cloud formation}
%\label{sect_clouds}
Condensation of up to 200 types of liquids and solids is included in
the atmospheric equation of state. The formation and sedimentation of
clouds and depletion of condensible material is treated in the
self-consistent timescale approach of \ct{Allard12}.
%The present set of atmosphere models is using modifications of the convection treatment described in \S\ref{sect_convection} to estimate the convective mixing, overshoot and gravity wave timescales that are at the core of this model. 
They used a monodisperse grain size distribution for
each cloud layer, but current version of the models now considers a log-normal distribution
over 1.8 decades in size in 12 bins with the altitude-dependent 
size determined by the same timescale formalism as in \ct{Allard12}. The set of 60
grain types included in calculating the cloud opacity has also been
updated with the optical data for several low-temperature condensates,
which become  important in the atmospheres of T dwarfs,
which are not discussed in this paper (Homeier et al., in prep).

\subsection{Solar abundances}
 Solar abundances have been revised over the past decade based on
 radiation hydrodynamical simulations of the solar photosphere
 combined with 3D NLTE radiative transfer \cp{Asplund09, Caffau11}. This
 leads to a substantial reduction of carbon, nitrogen, and oxygen
 abundances compared to the abundances of \ct{Grevesse93}, previously
 used in the NextGen and AMES-Dusty/Cond models. 
The present models
 adopt the solar composition of \ct{Asplund09} with revisions of the
 elemental abundances of C, N, O, Ne, P, S, K, Fe, Eu, Hf, Os, and Th
 obtained by the CIFIST project \cp{Caffau11}. This yields, in particular, a slight upward revision of carbon, nitrogen, and
 oxygen abundances and an increased total heavy element fraction by
 mass, with  $Z=0.0153$ compared to 0.0122 in \ct{Asplund05} and
 0.0134 in \ct{Asplund09}. Although the individual differences with the
 \ct{Asplund09} elemental abundances are modest, \ct{Antia11} and \ct{Basu13}
 have found that the higher overall metal content of the CIFIST
 abundances is in  better agreement with
 helioseismology results for the solar interior. 
 %To use a consistent
 %composition for the early phases of stellar evolution we are also
 %using the protosolar helium abundance of 10.98
% ($Y_\mathrm{ini}=0.271$) following \ct{Serenelli10} and \ct{Basu13}.
The helium abundance adopted in the atmosphere models is fixed to $Y_\mathrm{ini}=0.271$, which is representative of the initial solar helium abundance \cp{Serenelli10,Basu13}.

\begin{figure}
\vspace{-1.5cm}
 % \resizebox{\hsize}{!}{\includegraphics{figure1.eps
 % \resizebox{\hsize}{!}{\includegraphics[scale=0.4]{fig_nabla1.pdf}
 % \includegraphics[scale=0.48]{pt_fig.pdf}
  \includegraphics[height=16cm,width=10cm]{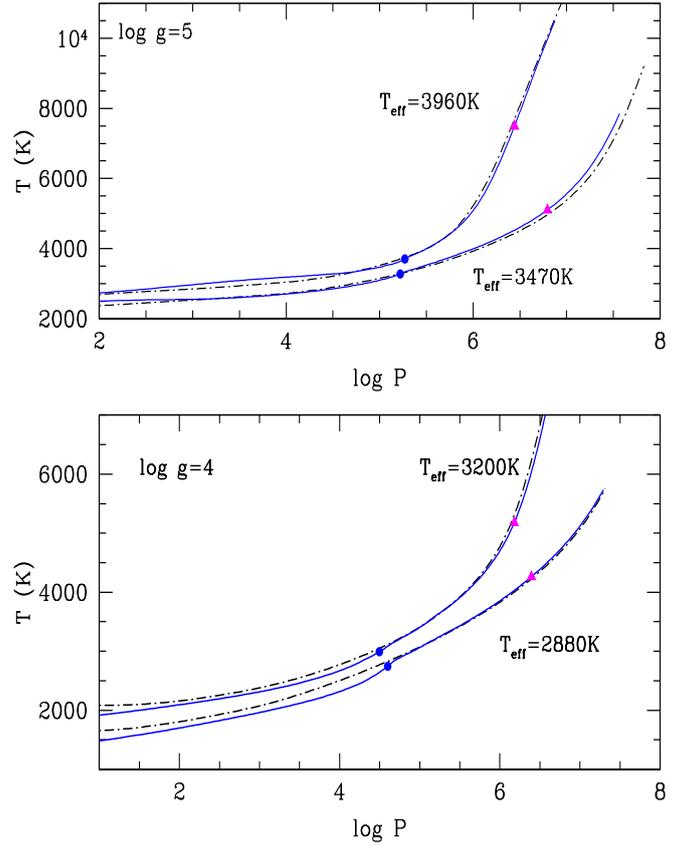}
  \vspace{-2.5cm}
   \caption{Comparisons of pressure-temperature atmospheric profiles between present 1D (solid blue) and CO$^5$BOLD RHD (dash-dot black) models, for different effective temperatures and surface gravities,
   as indicated in the panels for each curves. Pressure unit is dyne/cm$^{-2}$. Calibration of $\lmix$  based on such comparisons yields $\latm = 1.64 \times \Hp$ for $\te=2880,\logg=4.0$; $\latm = 1.58 \times \Hp$ for $\te=3200, \logg=4.0$; $\latm = 1.69 \times \Hp$ for $\te=3470, \logg=5.0$; $\latm = 1.58 \times \Hp$ for $\te=3960,\logg=5.0$.  The dot on each solid curve indicates the onset of convection and the triangle indicates the location of optical depth $\tau$=100 \cp{Chabrier97} where the transition to the interior model is performed.}
  \label{fig_PT}
\end{figure}

\subsection{Treatment of convection}
\label{sect_convection}
 Radiative-convective equilibrium in the \texttt{PHOENIX} atmosphere models is calculated using the mixing length theory \cp[or MLT, see][]{KW1990}. The main free parameter within this framework is the mixing length $\lmix$, expressed in terms of the pressure scale height $\Hp$ and
 $\lmix$ describes the efficiency of the convective energy transport in terms of the superadiabaticity of the temperature gradient. 
% The main changes concern choices for $\lmix$ and $\Gamma$ and are summarised in Table \ref{table1}. 
All previous \texttt{PHOENIX} models have used a constant value for $\lmix \equiv \latm$:  $\latm = \Hp$ in the NextGen and Cond/Dusty models and $\latm = 2 \times \Hp$ in \ct{Allard12}.

In the present work, the  determination of $\latm$ is inspired by the work of \ct{Ludwig99,Ludwig02}. It is based on comparisons between 1D MLT models and 2D/3D RHD simulations that cover main sequence and pre-MS
models down to the hydrogen-burning limit and below \cp{Freytag10, Freytag12}. Since the RHD
models include a basic treatment of dust formation, they also allow the
 calibration of late M dwarfs, where cloud opacity is becoming
relevant. This calibration yields  a value of $\latm\approx 1.6 \times \Hp$  for the Sun. The value of $\latm$ increases for later type stars, up to values $\simgt$ 2$\times \Hp$ for the coolest and densest models ($\te < 3000$ K and $\log g >$ 4.5). Comparisons between 1D and RHD pressure-temperature profiles are shown in Fig.~\ref{fig_PT}. The adopted value of $\latm$, for a given $\te$ and $\log \, g$, provides the best overall agreement that can be obtained; however, the agreement is not perfect all the way from the top to the bottom of the atmosphere,  reflecting limitations in the MLT formalism for properly handling atmospheric convection (Homeier et al., in prep).

\subsection{Evolutionary models}
\label{sect_grid}

\begin{figure}
 \includegraphics[height=12cm,width=9cm]{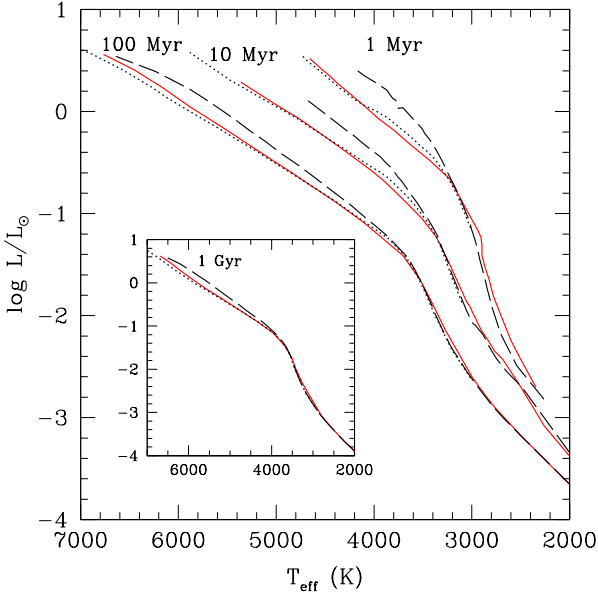}
\vspace{-2cm}
  \caption{Comparison of present models with the BCAH98 models for various isochrones, as indicated in the figure and inset.  Solid (red):  present models; Long dash (black): BCAH98 models with $\lint=  \Hp$; Dotted (black): BCAH98 with $\lint= 1.9 \times \Hp$%, which fits the Sun at its age
  .}
%:
  \label{fig_hrd2}
\end{figure}

%We have run evolutionary models for a mass range similar to the BCAH98 models, from 0.01 $\msol$ to 1.4 $\msol$. 
Current models cover the evolution of pre-MS and MS stars from 0.07 $\msol$ to 1.4 $\msol$.
Calibration of a 1 $\msol$ star sequence to fit the fundamental parameters of the Sun ($R_\odot$, $L_\odot$) at its age requires  a value  for the \emph{{\it
\emph{interior structure}}} of the mixing length $\lmix \equiv \lmix^{\rm int} = 1.6 \times \Hp$ and a helium abundance $Y=0.28$. We used this value of helium for the complete grid of evolutionary models. The value adopted in the atmosphere models is slightly different ($Y=0.271$, see \S 2.2). To appreciate the effect of this inconsistency, we have recalculated a full grid of evolutionary models assuming the same He abundance $Y$=0.271 in the interior structure as in the atmosphere models.  The effect of such a variation in Y on $\te$ for the whole range of masses considered is less than 1\%. The maximum effect on the luminosity is found for the highest masses ($M \simgt 1 \msol$) and reaches at most 10\%. For a 1 $\msol$ star, the effect on the luminosity is less than 5\% for ages $\simlt$ 2 Gyr, and up to 10\% for ages $\simgt$ 2 Gyr. We have also computed a few atmosphere models again with $Y=0.28$ for a range of $\te$ between 2600 K and 6000 K and of gravities $\log  g$ between 3.5 and 5. The effect of such a variation in Y is negligible on the atmospheric thermal profiles  (less than a 1\% effect on the pressure at a given temperature) and on the photometry (differences less than 0.01 mag for all colours explored).
  
We fixed $\lmix^{\rm int} = 1.6\times \Hp$ for the complete grid of evolutionary models for  the interior structure.
A fully consistent approach would be to adopt the value obtained for the corresponding atmosphere model, i.e  $ \lint=\latm$. We checked that this is unnecessary.
Indeed, values of  $\latm$  only start to depart significantly from $1.6\times \Hp$ for the coolest and  densest models ($\te < 3000$ K and $\log g >$ 4.5), reaching  values up to 2$\times \Hp$ or more. This concerns masses $< 0.2 \msol$, where a variation of $ \lmix^{\rm int}$ between 1.6$\times \Hp$ and 2$\times \Hp$  has an insignificant effect on the evolution. 

%since a value of $\lint =1.6 \times \Hp$ is closed to the RHD values of  $\latm$ corresponding to the mass and age range where evolutionary models are sensitive to the choice of $\lmix^{\rm int}$  \cp[see][]{Chabrier97, Baraffe02}. Values of  $\latm$  start to depart significantly from our canonical interior value only for the coolest and  densest models ($\te < 3000$ K and $\log g >$ 4.5), reaching  values $\simgt$ 2$\times \Hp$. This concerns masses $< 0.3 \msol$, where variation of $ \lmix^{\rm int}$ between 1.6$\times \Hp$ and 2$\times \Hp$  has insignificant effect on the evolution.
 
%We have also computed models for $ \lmix^{\rm int}  = 2 \times \Hp$, for the sake of comparison.
% Note that for masses typically $\le 0.6 \msol$, variation of $ \lmix^{\rm int}$ between 1.6$\times \Hp$ and 2$\times \Hp$, the later value being the one used in the atmosphere models for M dwarfs, has unsignificant effect on the evolutionary models (see Fig. \ref{fig_hrd1}). 
%Our new treatment of convection in the optically thin layers (i.e new MLT) has a noticeable effect on isochrones for low gravities, as mentioned in \S \ref{sect_convection}, corresponding to ages $\simlt 10 $ Myr. 
As illustrated in Fig. \ref{fig_hrd2}, the new models predict different positions of isochrones  than do the BCAH98 models. 
%, with in particular  slightly hotter $\te$ for a given luminosity and age. This is partly due to the increase of the value of  $ \lmix$ in the present atmosphere models, whereas $ \lmix= \Hp$ was used in the NextGen atmosphere models. 
Below $\te \simlt 2800\,$K, dust starts to form in the upper atmosphere and affects atmospheric thermal profiles and spectra. %We are still working on models including dust formation and settling, the so-called BT-settl models \cp{Allard12}, which 
 In the current grid of evolutionary models with 
  $\te > 2000$\,K, the effects of cloud opacity only gradually become
  visible for the latest M and early L dwarfs ($\te \simlt 2500\,$K).
  %and settling are not yet important. 
  They are becoming crucial for models of L and T dwarfs, which will be examined in a forthcoming study.
% late M and the earliest L really are included already with the
% tracks down to 2000 K.

\begin{figure}[htp!]
\includegraphics[scale=0.45]{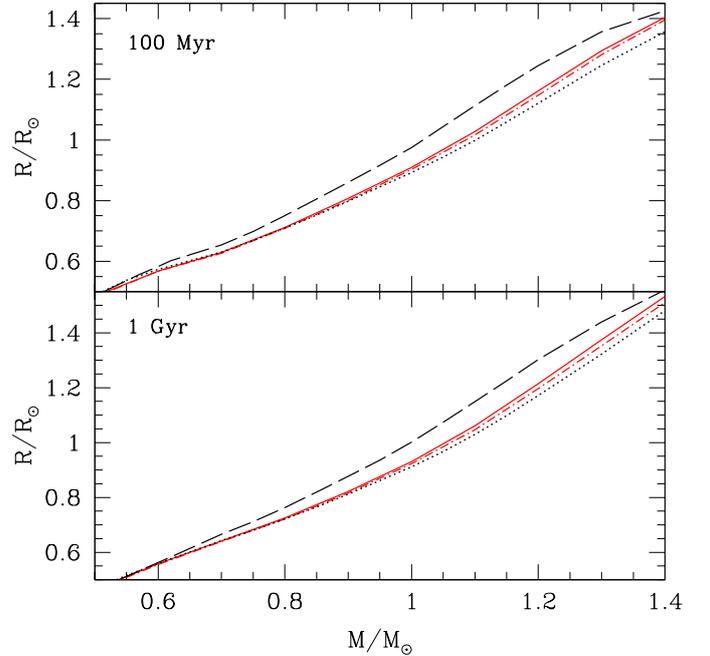}
\vspace{-2cm}
  \caption{Comparison of theoretical mass-radius relationships for ages of 0.1 Gyr (upper panel) and 1 Gyr (lower panel) and based on  different sets of models: Present models with $Y_{\rm int}=0.28$ (solid red) and $Y_{\rm int}=0.271$ (short-dash red); BCAH98 models with $\lint=  \Hp$ (long-dash black) and $\lint=  1.9 \times \Hp$ (dot black).}
%:
  \label{fig_mr}
\end{figure}

Theoretical mass-radius relationships for 0.1 Gyr and 1 Gyr are shown in Fig. \ref{fig_mr} for different sets of models. As shown in \ct{Chabrier97}, the mass-radius relationship for low-mass stars is essentially fixed by the equation of state, with a very small dependence on the atmosphere treatment. The differences between present models and the BCAH98 models essentially stem from the different values of the interior mixing length $\lint$ used. As shown in Fig. \ref{fig_mr}, the effect of the interior He abundance $Y_{\rm int}$ slightly affects the mass-radius relationship for masses $\simgt$ 1 $\msun$.

\vspace{-1cm}

\begin{figure}[htp!]
\includegraphics[scale=0.45]{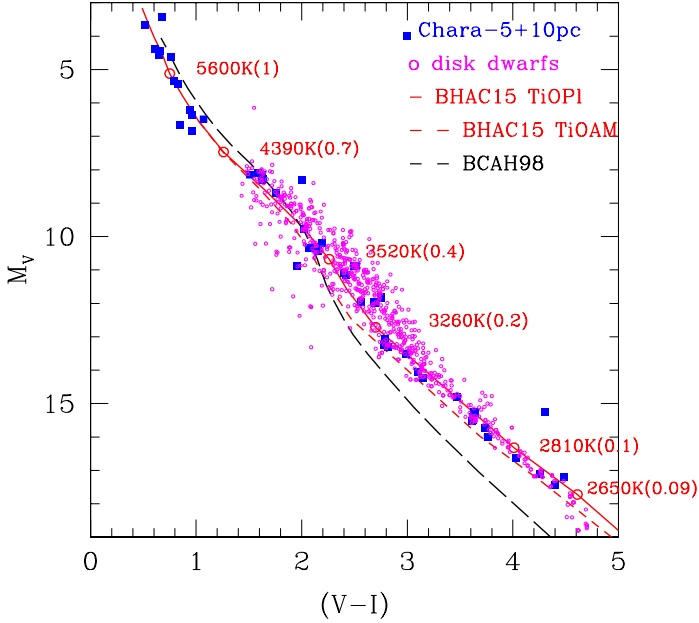}
\vspace{-2cm}
  \caption{Comparison of 1 Gyr isochrones with disk dwarfs in an optical CMD.   Present models based on the Plez TiO  (solid red) and the AMES TiO (dash red) linelists are shown, along with the 
 BCAH98 models (long dash black). 
  %; Dash dot (black):  set of models following the BCAH98 with AMES TiO. 
  The magenta open circles are
  disk objects with parallax \cp{Monet92, Leggett92} and the blue
  squares are from the recent 5\,pc and 10\,pc solar neighbourhood samples
  of \ct{Cantrell13} with more accurate parallaxes and cleaned of
  binary systems. The numbers close to the red open circles on the
  solid curve give $\te$ and mass (in $\msol$ in the brackets) for
  selected models.}
%:
  \label{fig_magviv}
\end{figure}

\section{Comparison with observations}

In this section, we highlight the significant improvement provided by this new set of models, which solve some of the  problems of the BCAH98 models, through a selection of observational tests. %We will thus not present exhaustive comparisons with observations. 
Colour-magnitude diagrams (CMDs) provide the best validity test, as they test the consistency between interior and atmosphere structures.

\subsection{Optical colours}
\label{sect_optical}

As mentioned in the introduction, a well-known flaw of the BCAH98 models is the prediction of colours that are too blue ($V\!-\!I$). Figure \ref{fig_magviv} compares models at 1 Gyr with solar-neighbourhood and disk objects with parallax \cp{Monet92, Leggett92, Cantrell13}. The comparison shows that this persistent problem of the models seems to be solved by several improvements. The most important one is the use of the TiO linelist from \ct{plezTiO}. The impact of the TiO {linelist has already been noticed with the generation of models following the NextGen ones, such as the Dusty models \cp{Chabrier00b, Allard01}. The former ones use the linelist by \ct{Jorgensen94}, while the latter  ones include the AMES TiO linelist of \ct{Schwenke98}. The improvement in ($V\!-\!I$) colours of models that include the AMES TiO linelist of \ct{Schwenke98}  was significant (see Fig. \ref{fig_magviv}), but still unable to yield satisfactory agreement with observations. Indeed, an offset in ($V\!-\!I$) colours of $\sim 0.4$ mag  remained between $\te ~ \sim$ 3600\,K  ($\mv \sim$ 10) and $\te ~ \sim$ 2300\,K   \cp[$\mv \sim$ 19, see][]{Chabrier00b}.
Improvement due to the implementation of the \ct{plezTiO} TiO 
%\textbf{and VO}
%% May need additional discussion of the effects of VO (more flux in $V$)
%% and perhaps of MgH (only in $B$...)  DH
linelist in the new models is illustrated in
Fig. \ref{fig_magviv} by the comparison between present models
(TiOPl) and the same models  based on the AMES TiO
linelist (TiOAM).
%despite remaining shortcomings in the opacity data (cf.\ \S \ref{sect_lines}).

Another positive effect stems from the change in solar abundances,
with a decrease in oxygen abundance by $\sim$22\% between the
\ct{Grevesse93}, used in previous sets of models, and the presently used \ct{Caffau11}
abundances. 
%The new Ti and V solar abundances  are also slightly decreased compared to the \ct{Grevesse93} abundances, but only by $\sim$ 14\%-15\%.  Despite this decrease of Ti and V, translating in a decrease of the mixing ratios of TiO and VO by about the same amount, the dominant effect is provided by the larger decrease of the O abundance and consequently of H$_2$O.  Indeed, as oxygen binds preferentially to many of the metals present, including Ti and V, this yields an even more significant decrease of the total water mixing ratio  by 30\%-35\%, compared to that of TiO and VO. 
As a consequence of relatively less O, the flux in the IR generally increases owing to weaker water absorption.  Because of flux redistribution, the flux in the $V$-band decreases relative to the flux at longer wavelengths. This is an important  effect, illustrating how water abundance affects optical colours due to flux redistribution. 
  
 % We  still note, however, a small offset of the predicted colours for the new models, remaining slightly too blue compared to observations. The reason for this discrepancy is not clear. Further tests and experiments with convection treatment in the atmosphere models did not provide better agreement. We thus think that this stems from some remaining uncertainties in opacities, that are however hard to identify (i.e optical or near-IR absorber) given the small effect required and possible unexpected effects due to flux redistribution as above mentioned with water.
  
\subsection{Near-infrared colours}
\label{sect_nearIR} 

\begin{figure}
%\vspace{-1cm}
%\hspace{-1cm}
%\resizebox{\hsize}{!}{\includegraphics{fig_magiki.pdf}{fig_magjhj.pdf}
%\epsscale{1.}
%\hspace{1.8cm}
 %\includegraphics[scale=0.35]{fig_magiki.pdf}
% \includegraphics[scale=0.35]{fig_magjhj.pdf}
%\includegraphics[scale=0.45]{fig_magiki-jhj.pdf}
\includegraphics[width=9cm, height=14cm]{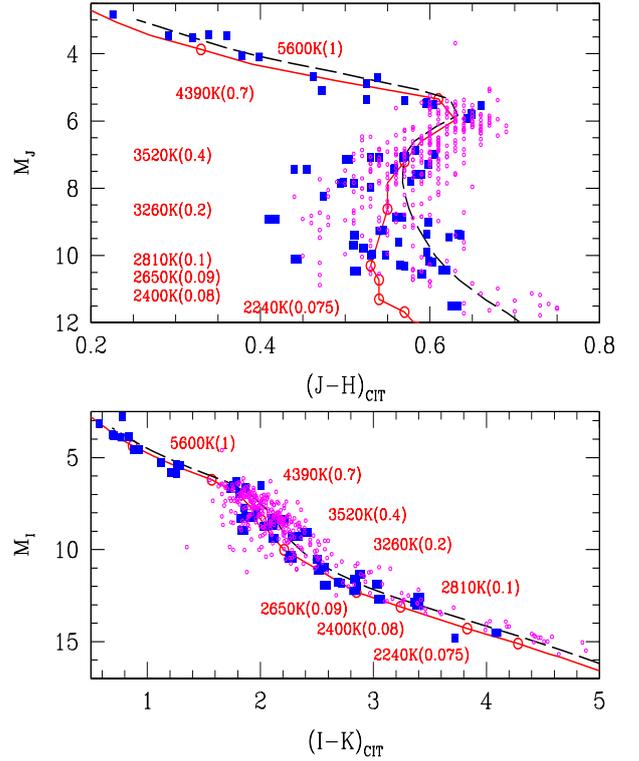}
\vspace{-2.5cm}
  \caption{Comparison of 1 Gyr isochrones with disk dwarfs in near-IR CMDs in CIT filters.   Solid (red):  new models; Long dash (black): BCAH98 models with $\lint=  \Hp$. Symbols are the same as in Fig. \ref{fig_magviv}.}
%:
  \label{fig_IR}
\end{figure}

The new models yield also a significantly better match with observations for
relatively old objects (age$\simgt$1 Gyr, $\log g \simgt
4.5$) in near-IR CMDs, as illustrated in Fig. \ref{fig_IR}. This stems from the
use of the more complete and accurate water linelist of \ct{Barber06}, compared to previously
used ones, namely  \ct{Miller94} in NextGen and  \ct{Partridge97} in Dusty/Cond.

%Despite, indeed, improvement in optical colours of models including an improved TiO linelist compared to that used in the NextGen models, the AMES linelist for water \cp{Partridge97} as used in the Dusty models, had worsened the agreement with observations of M dwarfs (for $\te \simgt$ 2300\,K) in near-IR colours, as discussed in \ct{Chabrier00b}.
%%This known flaw of models following the NextGen models prevented an earlier release of a successor grid of the BCAH98 models. 
%%with the main molecular libelists used in the Dusty models.  
%%While the NextGen models used the \ct{Miller94} linelist, the Dusty models used the AMES linelist from  \ct{Partridge97}, and the new models is based on the much more complete \ct{Barber06} linelist. 
%Unexpectedly enough, although the \ct{Miller94} water linelist used in the NextGen models was less complete for the higher energy transitions than the AMES linelist, the former yielded better agreement between models and IR photometric observations for $\te \simgt$ 2300\,K, as it relied on better potential surfaces for high temperatures \cp{Allard01}. 
%The more complete and accurate water linelist of \ct{Barber06} used in the present models solves this problem.

To illustrate the impact of the new models at low gravities and in a similar range of effective temperatures ($\te > 2000$K),
%the new MLT treatment in the atmosphere models (see \S \ref{sect_convection}), 
we also compared them to data in the young cluster $\sigma$ Orionis \cp{Pena12}, with an age of a few Myr ($\log g \simlt 4$). The results are shown in Fig. \ref{fig_ori} in various CMDs. Fluxes from the models have been transformed in the VISTA filters to be consistent with the data of \ct{Pena12}. Comparison of models in CMDs (or luminosity-$\te$ diagrams) with observations in 
 young clusters and star formation regions should always be taken with caution owing to the observed luminosity spread  and the effect of initial conditions and accretion history \cp{Baraffe02, Baraffe09, Baraffe12}. The comparison in Fig. \ref{fig_ori}, however, is illustrative of the overall improvement in $Z\!J\!H\!K$ filters of the new models compared to the BCAH98 ones. The latter indeed provide poor agreement with observations at young ages in near-IR CMDs, as illustrated in Fig. \ref{fig_ori}. 
 For this reason, models had to be compared with data in $L-\te$ or HR diagrams, which implied the use of very uncertain bolometric corrections and spectral type-$\te$ or colour-$\te$ scales.
 The agreement of the new models with observations, however, deteriorates at the faintest magnitudes by up to $\sim$ 0.1 mag in  ($J\!-\!H$) and $\sim$ 0.4 mag in  ($J\!-\!K$).
% , with models predicting slightly too blue ($J\!-\!K$) colours ($\sim$ 0.1 mag discrepancy). 
 Given photometric uncertainties and possible $K$-band excess at this age owing to the presence of a circumstellar disk, it is premature to draw any conclusion on the reason for these discrepancies.
  
 %Figures \ref{fig_ori} also show differences in the positions of isochrones between the standard and the "newMLT" treatments of convection in the atmosphere. Though the latter treatment is probably physically more sound, it is hard to tell from such comparisons whether it provides a better agreement with observations, given the spread of the data, the distance and age uncertainties of the cluster and extinction effects. 
 %Further comparisons with data in young clusters may help discriminate between those two sets of models. 
 
  \begin{figure}
%\resizebox{\hsize}{!}{\includegraphics{fig_magiki.pdf}{fig_magjhj.pdf}
%\epsscale{1.}
%\hspace{-0.5cm}
% \includegraphics[scale=0.30]{fig_orizjj.pdf}
% \includegraphics[scale=0.30]{fig_orijhj.pdf}
 %\includegraphics[scale=0.30]{fig_orijkj.pdf}
  \includegraphics[scale=0.40]{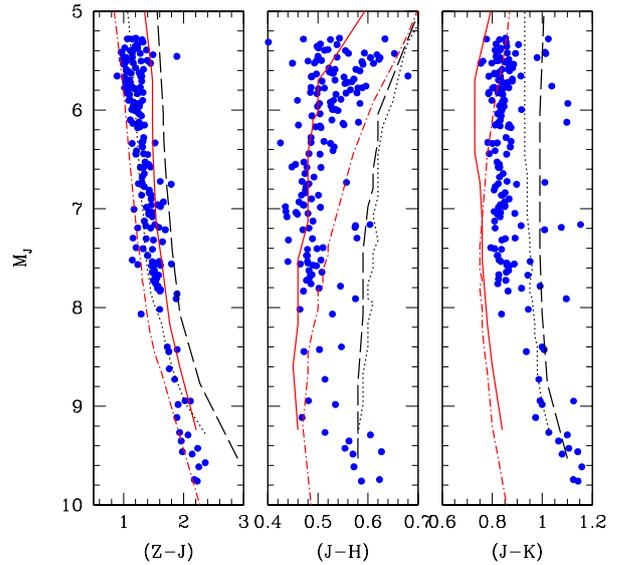}
  \vspace{-2cm}
  \caption{Comparison of  models with observations in $\sigma$ Orionis in various CMDs (VISTA filters). Isochrones of 1 Myr and 10 Myr are displayed for various sets of models:
Present models (red): solid (1 Myr) and dash-dot (10 Myr).
 BCAH98 models  with $\lint=  \Hp$ (black): long dashed (1 Myr) and dot (10 Myr). The data (blue dots) are from \ct{Pena12} using a distance modulus of 7.73 (i.e. a distance of 352 pc). The hydrogen-burning limit ($M$=0.07 $\msol$)  is located at $M_{\rm J}\sim$ 6 ($\te \sim 2900$K, $\log g \sim$ 3.5) at 1 Myr and $M_{\rm J}\sim$ 8 ($\te \sim 2950$K, $\log g \sim$ 4.2) at 10 Myr.}
%:
  \label{fig_ori}
\end{figure}

 \subsection{Multiple systems}
\label{sect_multiple} 

Stringent tests of evolutionary models are provided by coeval multiple systems. The recent discovery of the quadruple pre-main sequence system LkCa 3 is interesting  in this respect  \cp{Torres13}. If one admits that the four components of LkCa 3 should be coeval and must  lie on the same isochrone, they make an excellent test of evolutionary models. This supposes  that early history of accretion has not altered their structure \cp{Baraffe09, Baraffe12} and that their position in a CMD or $L-\te$ diagram can be reproduced by non-accreting evolutionary models.  

With this assumption, \ct{Torres13} find that no currently available model is able to reproduce the position of the four components with one single isochrone in a $(V\!-\!H) - M_{\rm V}$ CMD. In contrast,  our new models do fulfil this stringent constraint within the error bars (see Fig. \ref{fig_lkca3}). The new models agree with observations for an age of 1.6 Myr. In comparison, the best-fit 6.3 Myr isochrone of the BCAH98 models with $\lint=  \Hp$ is unable to match the coolest component LkCa 3 Ab, which can be fitted by an isochrone of  2.5 Myr. A similar problem arises with the BCAH98 models using $\lint=  1.9 \times \Hp$. 
The same failure was reported by \ct{Torres13} (see their Fig.~6) with the Dartmouth models \cp{Dotter08}. 
%Here again, however, it is difficult to discriminate between the standard and new MLT models, as they can both reproduce the four components, but for a different age. 

As a final note, even if accretion history can have an impact on the structure of young objects, as suggested by \ct{Baraffe12}, this does not necessarily affect all objects in the same way. Even for binary or multiple systems, each component has its own accretion history,  and its structure may
be differently altered by accretion, since the effect depends on the total amount of  accreted mass (see Baraffe et al. 2012 for details). Disentangling the remaining uncertainties of models and possible effects of accretion when comparing models to young multiple systems is thus not straightforward. 
 We hope that many more such multiple systems, spanning a wide range of masses and ages, will be discovered in the near future to provide more clues to the remaining model uncertainties (e.g. convection treatment, opacity uncertainty) and/or to the impact of accretion for the youngest systems. 

 \begin{figure}
%\resizebox{\hsize}{!}{\includegraphics{fig_magiki.pdf}{fig_magjhj.pdf}
%\epsscale{1.}
\vspace{-1cm}
 \includegraphics[width=9cm, height=12cm]{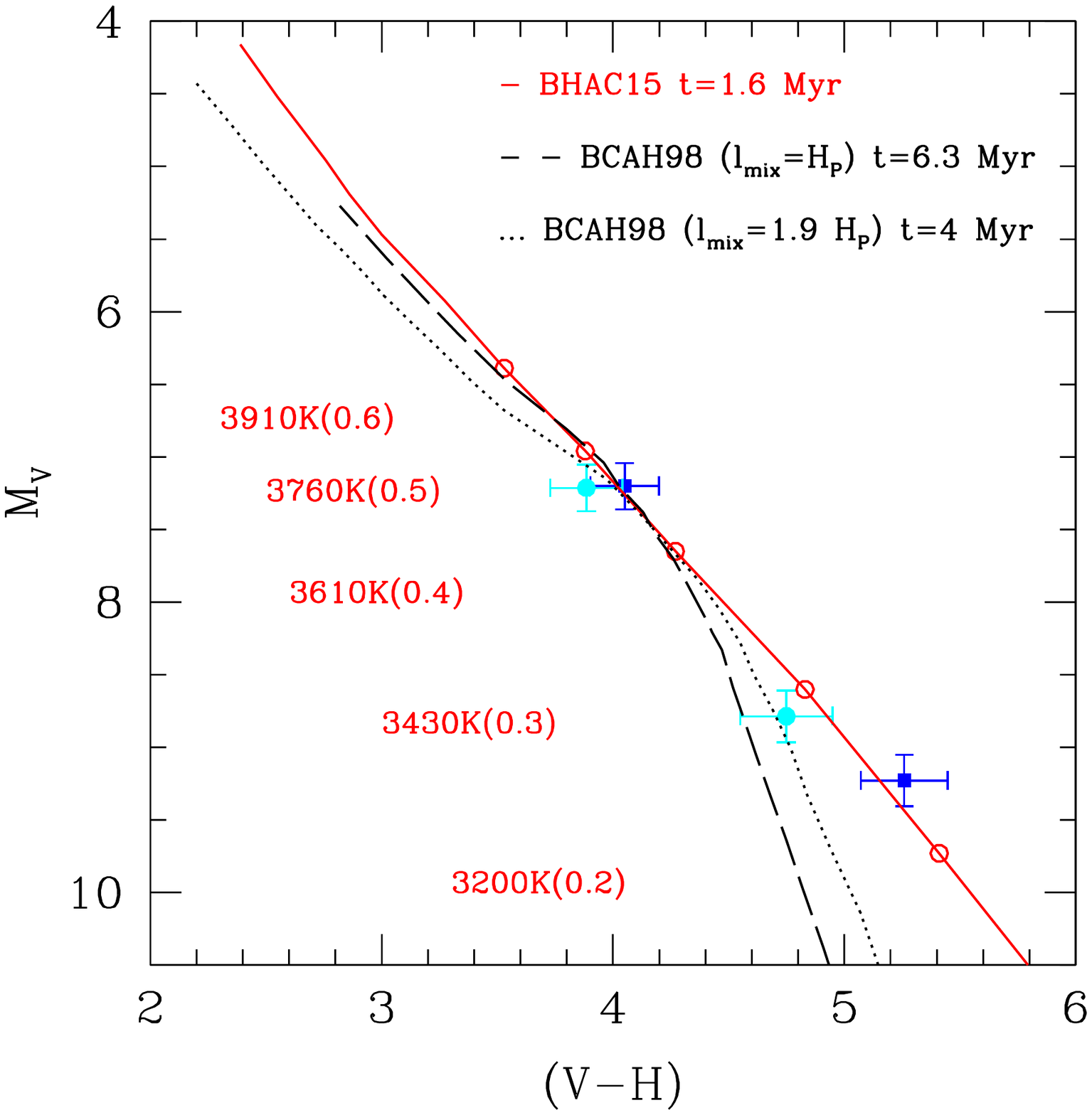}
 \vspace{-2cm}
  \caption{Comparison of  models with the four components of the quadruple system LkCa 3 \cp{Torres13}. The best-fit isochrone is shown for various sets of models. 
Present models for an age of 1.6 Myr (solid red).  BCAH98 models with $\lint=  \Hp$ for an age of 6.3 Myr (long-dash black). BCAH98 models with $\lint=  1.9 \times \Hp$ for an age of 4 Myr (dot black). The blue dots are for  LkCa 3\,A and the cyan squares for LkCa 3\,B. We adopt a distance $d=127$ pc and extinction $E(B-V) = 0.10$ as in \ct{Torres13}. The numbers close to the red open circles on the solid curve give $\te$ and mass (in $\msol$ in the brackets) for a few models.}
%:
  \label{fig_lkca3}
\end{figure}

\subsection{Tests of stellar evolutionary models}

%Finally, let us stress  the importance to use  models consistently coupling internal and atmospheric structures when 
For the sake of testing evolutionary models against observations,
 it is crucial to use the synthetic colours and magnitudes predicted by the {\it \emph{same}} atmosphere models as those that provide the outer boundary conditions for the interior structure. Using  ``hybrid'' models, based on atmospheric thermal profiles for the interior structure from one set of atmosphere models and colours/magnitudes from another set,  yields incorrect results. Figure \ref{fig_consistent} illustrates this point, showing in two CMDs for various isochrones a comparison between the present set of models, the BCAH98 models, and a hybrid model
based on BCAH98 evolutionary models (relying on the NextGen atmosphere models) and the colours and magnitudes from current atmosphere models. The positions and slopes of the hybrid model isochrones are significantly different from both the new and the BCAH98 models. Using these and similar hybrid models to derive cluster ages and masses or to test them against multiple systems yields incorrect results and incorrect conclusions on the validity or invalidity of the models.
 Such comparisons, often seen  in the literature, are totally meaningless.
%Internal structure models and atmosphere models cannot be decoupled as they share the same thermal profile which is crucial for both the thermal balance of the interior structure and the synthetic spectrum. 
 
 \begin{figure}
 \includegraphics[width=9cm, height=13cm]{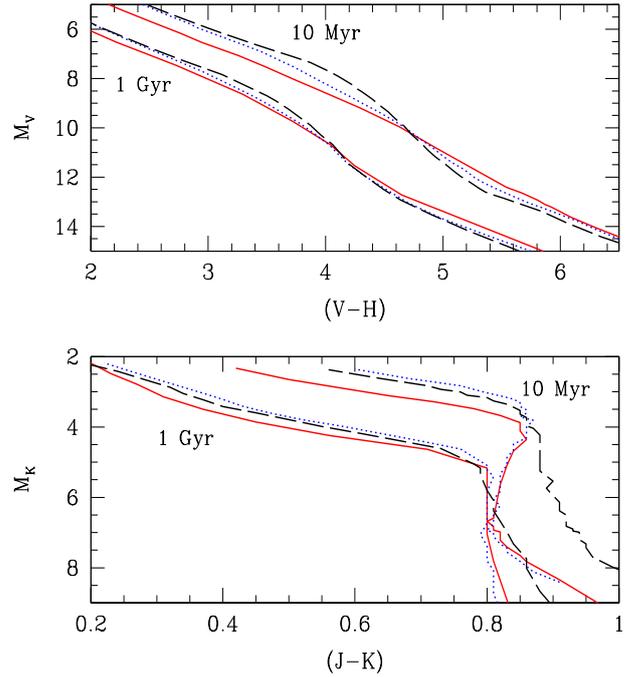}
  \vspace{-2cm}
  \caption{Comparison of 10 Myr and 1 Gyr isochrones for various sets of models in two CMDs. 
Solid (red): Present models. 
Long dash (black): BCAH98 models  with $\lint=  \Hp$. Dot (blue): hybrid models using the BCAH98 structure models and the colours/magnitudes of present atmosphere models. }
  \label{fig_consistent}
\end{figure}

\section{Conclusions}

The new set of evolutionary models presented in this paper, consistently coupling internal and atmospheric structures,
show  significant improvement over previous generations. They solve some of the persistent flaws present in them, such as predicting optical colours that are too blue. Significant improvement in the atmosphere models, in terms of updated molecular linelists and revised solar abundances, provide a much better match to observations. 
%We still note some remaining discrepancies, namely predicted ($V\!-\!I$) colours which are still slightly to/o blue for a given $M_{\rm V}$.  Identifying the culprit is  not straightforward, given the small effect required to bring models and observations into agreement. 
%It may not necessarily stem from missing opacities in the optical, since as illustrated for water vapour, a slight alteration of flux absorption in the near-IR may as well affect the flux redistribution in the optical and impact optical colours as well. 
More systematic tests of current models against  observations will help to identify the remaining uncertainties. 
%Despite remaining uncertainties in our treatment of atmospheric convection within the framework of the MLT, we do not think that this can still be a source of large uncertainty, at least for relatively old (age $>$ 10 Myr), high gravity ($\log g > 4.5$) objects. 
The calibration of the mixing-length parameter, as presented in this work, provides overall good
agreement between 1D and multi-D RHD thermal profiles. This exercise highlights the poor approximation of using a constant value for $\latm$ for pre-main sequence and main-sequence low-mass stars.  
The agreement, however, is  not perfect, depending on the effective temperature and the surface gravity. This very likely reflects the limitation of the MLT formalism
to correctly capture the thermal properties of convectively unstable atmospheres over a wide range of parameters.  
 We note as well that remaining uncertainties in the multi-D RHD simulations are not excluded. 
  Efforts towards developing such
simulations in cool atmospheres need to be pursued. Since models at
young ages are particularly sensitive to the treatment of atmospheric
convection \cp{Baraffe02}, comparisons between models and observations  of objects in
young clusters (age $\simlt$ 10 Myr)
will help to improve  the treatment of
convection.

Following the release of these new models down to the hydrogen-burning limit \footnote{Isochrones are available at:

http://emps.exeter.ac.uk/physics-astronomy/staff/ib233 

or http://perso.ens-lyon.fr/isabelle.baraffe/BHAC15dir}, we are currently
working on their extension to L, T, and Y dwarfs, including dust formation and settling. This will provide a complete and coherent set of models for low-mass stars and brown dwarfs down to Jupiter-mass objects.

\section{Acknowledgements}
This work is supported by Royal Society award WM090065. It is also
partly supported by the ERC under the European
Community's Seventh Framework Programme (FP7/2007-2013 Grant Agreement
No. 247060),  the consolidated STFC grant ST/J001627/1, the
French ANR through the GuEPARD
project grant ANR10-BLANC0504-01, and the PNPS of the CNRS (INSU).
We thank the CHARA team (Russel White, Todd Henry, Cassy Davidson, Justin Cantrell) for providing the CHARA data and Guillermo Torres for providing information on LkCa 3. We thank Jos{\'e} Caballero for discussion and Richard Freedman and Jonathan Tennyson for sharing their insights into the status of molecular line data. 
The model atmospheres used in this study were calculated at the
PSMN of the
ENS-Lyon and at the GWDG in
co-operation with the Institut f{\"u}r Astrophysik G{\"o}ttingen. 
%We thank these institutions for generous allocation of computing time.
%The model analysis and synthetic photometry has made extensive use of the Numpy and AstroPy Python packages, and we thank contributors to these software projects.

\bibliographystyle{aa.bst}
\bibliography{references}

\end{document}